\documentstyle[11pt,twoside,colloqOHP,epsfig]{article}
\newcommand{\be}{\begin{equation}}
\newcommand{\ee}{\end{equation}}
\newcommand{\bea}{\begin{eqnarray}}
\newcommand{\eea}{\end{eqnarray}}
\begin{document}
\title{On the relation between Hot-Jupiters and the Roche Limit}
\author{Eric B.\ Ford$^1$, Frederic A.\ Rasio$^2$}
\affil{$^1$Astronomy Department, UC Berkeley, 601 Campbell Hall,
        Berkeley, CA 94709, USA [eford@astro.berkeley.edu]}
\affil{$^2$Dept. of Physics \& Astronomy, Northwestern, 2145 Sheridan Road, 
        Evanston, IL 60208-834 [rasio@northwestern.edu] }
\begin{abstract}
 Many of the known extrasolar planets are ``hot Jupiters,'' 
 giant planets with orbital periods of just a few days. 
We use the observed distribution of hot Jupiters to constrain the
location of the ``inner edge'' and planet migration theory.  If we
assume the location of the inner edge is proportional to the Roche
limit, then we find that this edge is located near twice the Roche
limit, as expected if the planets were circularized from a highly
eccentric orbit. If confirmed, this result would place significant limits
on migration via slow inspiral. However, if we relax our
assumption for the slope of the inner edge, then the current sample of
hot Jupiters is not sufficient to provide a precise constraint
on both the location and power law index of the inner edge.
\end{abstract}
\section{Introduction}
Early radial velocity discoveries were interpreted as showing
a pile-up at an orbital period of three days, but recent transit surveys and very sensitive
radial velocity observations have discovered planets with even shorter
orbital periods. These discoveries suggest that the inner edge for hot
Jupiters is not defined by an orbital period, but rather by a tidal
limit which depends on both the semi-major axis and the planet-star
mass ratio (See Fig.\ 1).  This would arise naturally if the inner
edge were related to the Roche limit, the critical distance at which a
planet fills its Roche lobe.  The Roche limit, $a_R$ is is defined by
\be
R_P = 0.462 a_R \left(\frac{M_P}{M_*}\right)^{1/3}
\ee
where $R_P$ is the radius of the planet, $M_P$ is the mass of the
planet, and $M_*$ is the mass of the star.
\begin{figure}[pbth]
\begin{center}
\epsfig{file=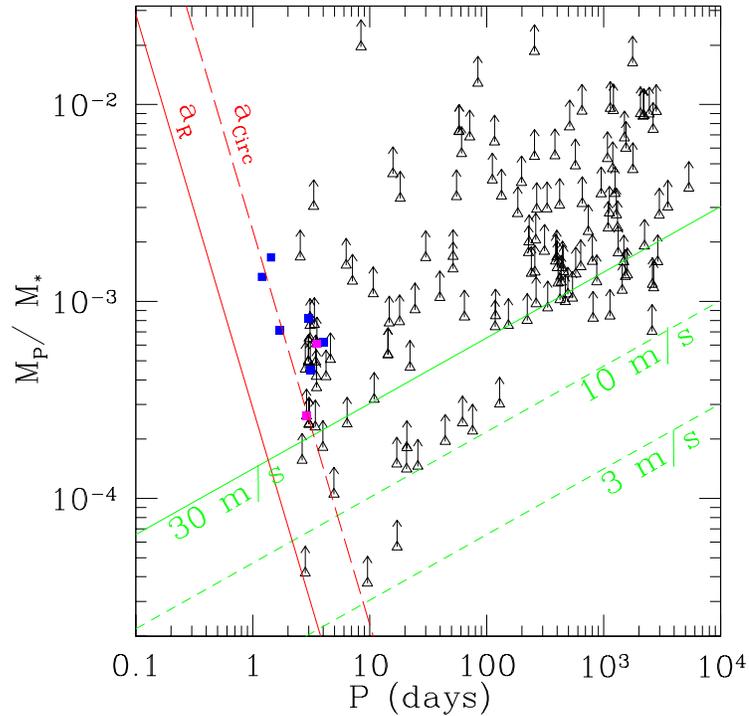,width=10cm}
\caption{
Minimum mass ratio versus orbital period for the current observed
sample of planetary companions.  Planets discovered by radial velocity
surveys are shown as triangles with arrows indicating $1-\sigma$
uncertainties in mass due to unknown inclination.  The magenta squares
have inclinations and radii measured via transits.  The blue squares
show planets discovered by transit searches.  The green lines show the
minimum mass corresponding to various velocity semi-amplitudes and
roughly indicate where radial velocity surveys are nearly complete
($\ge30$m/s), have significant sensitivity ($\ge10$m/s), and are only
beginning to detect planets ($\ge3$m/s).  The two red lines show the
the location of the Roche limit ($a_R$) and the ideal circularization
radius ($a_{\rm circ}$) for a planet with a radius, $R_P = 1.2 R_J$.
The inner edge for the distribution of hot Jupiters is near $a_{\rm
circ}$.  Note that the red lines do not apply to the lowest mass
planets that likely have a radius significantly less than $1.2 R_J$
due to their qualitatively different internal structure.
}
\end{center}
\end{figure}

The numerous mechanisms proposed to explain the migration of giant
planets to short period orbits can be divided into two broad
categories:
\begin{enumerate}
\item Mechanisms involving slow inspiral, such as migration due to a
gaseous disk or planetesimal scattering (Trilling \etal~1998, Gu \etal~2003).  
These would result in a limiting separation equal to the Roche limit.
\item Mechanisms involving the circularization of highly eccentric
orbits with small pericenter distances, possibility due to
planet-planet scattering (Rasio \& Ford 1996, Ford, Havlickova, \&
Rasio 2001, Papaloizou \& Terquem 2001, Marzari \& Weidenschilling
2002), secular perturbations from a wide binary companion (Holman,
Touma \& Tremaine 1997), or tidal-capture of free-floating planets
(Gaudi 2003).  These would result in a limiting separation of {\em
twice} the Roche limit (Faber \etal~2004).
\end{enumerate}

\section{Statistical Analysis}

To quantitatively explore the observational constraints on the
distribution of hot Jupiters, we employ the techniques of Bayesian
inference.  In the Bayesian framework, the model parameters are
treated as random variables which can be constrained by the actual
observations.  Therefore, to perform a Bayesian analysis it is
necessary to specify both the likelihood (the probability of making a
certain observation given a particular set of model parameters) and
the the prior (the {\em a priori} probability distribution for the model
parameters).  Let us denote the model parameters by $\theta$ and the
actual observational data by $d$, so that the joint probability
distribution for the observational data and the model parameters is
given by
\be
p(d, \theta) = p(\theta) p(d | \theta) = p(d) p(\theta | d), 
\ee
where we have expanded the joint probability distribution in two ways
and both are expressed as the product of a marginalized probability
distribution and a conditional probability distribution.  The prior is
given by $p(\theta)$ and the likelihood by $p(d|\theta)$.  On the far
right hand side, $p(d)$ is the a priori probability for observing the
values actually measured and $p(\theta|d)$ is the probability
distribution of primary interest, the {\em a posteriori} probability
distribution for the model parameters conditioned on the actual
observations, or simply the posterior.  The probability of the
observations $p(d)$ can be
obtained by marginalizing over the joint probability density and again
expanding the joint density as the product of the prior and the
likelihood.  This leads to Bayes' theorem, the primary tool for
Bayesian inference,
\be
p(\theta | d ) = \frac{p(d|\theta)p(\theta)}{p(d)} = \frac{p(d|\theta)p(\theta)}{\int \, d\theta p(d|\theta)p(\theta)}
\ee
Often the model parameters contain one or more parameters of
particular interest (e.g., the location of the inner cutoff for hot
Jupiters in our analysis) and other nuisance parameters which are
necessary to adequately describe the observations (e.g., the fraction
of stars with hot Jupiters in our analysis).  Since Bayes' theorem
provides a real probability distribution for the model parameters, we
can simply marginalize over the nuisance parameters to calculate a
marginalized posterior probability density, which will be the basis
for making inferences about the location of the inner cutoff for hot
Jupiters.

We construct models for the distribution of hot Jupiters and use
Bayes' theorem to calculate posterior probability distributions for
model parameters given the orbital parameters measured for
extrasolar planets discovered by radial velocity surveys.  For the
sake of clarity, we start by presenting a simplistic one-dimensional
model for the distribution of hot Jupiters.  We then gradually improve
our model to understand how each model improvement affects our
results.


The primary question which we wish to address in this paper is the
location of the inner edge of the distribution of hot Jupiters
relative to the location of the Roche limit.  Therefore, we define $x
\equiv a / a_R$, where $a$ is the semi-major axis of the planet and
$a_R$ is the Roche limit.  We assume that the actual distribution of $x$
for various hot Jupiters is given by a truncated power law,
\be
p(x | \gamma, x_l, x_u) dx = x^\gamma \left(\frac{dx}{x}\right), \quad x_l < x < x_u,
\ee
and zero else where.  Here $\gamma$ is the power law index and $x_l$
and $x_u$ are the lower and upper limits for $x$.  The lower limit,
$x_l$, is the model parameter of primary interest, while $\gamma$ and
$x_u$ are nuisance parameters.  Therefore, our results are summarized
by the marginalized posterior probability distribution for $x_l$.

In order to minimize complexities related to the analysis of a
population, we choose to restrict our analysis to a subset of the
known extrasolar planets for which radial velocity surveys are
complete and extremely unlikely to contain any false positives.  To
obtain such a sample, we impose two constraints: $P\le P_{\max}$, where
$P_{\max}$ is the maximum orbital period, and $K\ge K_{\min}$, where
$K_{\min}$ is the minimum velocity semi-amplitude.  We use
$K_{\min}=30$m/s, based on the results of simulated radial velocity
surveys (Cumming 2004).  We typically set $P_{\max}=$30d, even though
radial velocity surveys are likely to be complete even for longer
orbital periods (provided $K\ge K_{\min}$).  This minimizes the chance of
introducing biases due to survey incompleteness or possible structure in the
observed distribution of planet orbital periods at larger periods.  By
considering only planets with orbital parameters such that radial
velocity surveys are very nearly complete, our analysis does not
depend on the velocities of stars for which no planets have been
discovered.  Note that our criteria for including a planet may
introduce a bias depending on the actual mass-period distribution. 
We will address this point with a two-dimensional model at the end of
our analysis.
Also, our criteria exclude any planet
discovered via techniques other than radial velocities (e.g.,
transits), even if subsequent radial velocity observations were
obtained to confirm the planet. 

Initially, we make several simplifying assumptions to make an analytic
treatment possible.  We assume uniform prior probability distributions
for each of the model parameters, $p(\gamma) \sim
U(\gamma_{\min},\gamma_{\max})$ and $p(x_l,x_u) \sim \mathrm{const}$,
provided $x_{ll} < x_l < x_u < x_{uu}$ and zero otherwise.  The lower
and upper limits ($x_{ll}$ and $x_{uu}$) for each parameter are chosen
to be sufficiently far removed from regions of high likelihood that
the limits do not affect the results.
We assume that the orbital period ($P$), velocity semi-amplitude
($K$), semi-major axis ($a$), stellar mass ($M_*$), and planet mass
times the sin of the inclination of the orbit relative to the line of
sight ($m \sin i$) are known exactly based on the observations.  


We begin by assuming that $\sin i=1$ for all planets
and that all planets have the same radius, $R_P$.
With these assumptions, the posterior probability distribution is
given by 
\be
p(x_l, x_u, \gamma | x_1, ... x_n ) \sim \gamma^n (x_u^\gamma - x_l^\gamma)^{-n} 
\prod_{j=1}^n x_j^{\gamma-1},
\ee
provided that
$x_{ll} < x_l \le x_{(1)} \le  x_{(n)} \le x_u < x_{uu}$ and 
$\gamma_{\min} < \gamma < \gamma_{\max}$.  Here
$n$ is the number of planets included
in the analysis, $x_{(1)}$ is the smallest value of $x$ among the
sample of hot Jupiters used in the analysis, and $x_{(n)}$ is the largest
value of $x$ in the sample.  The normalization can be obtained by
integrating over all allowed values of $x_l$, $x_u$, and $\gamma$.  

We show the marginal posterior distributions in which we have
integrated over the nuisance parameters, $x_u$ and $\gamma$ in Fig.\ 2
(left, dotted line), assuming $R_P = 1.2 R_J$.  The distribution has a sharp
cutoff at $x_{(1)}$ and a tail to lower values reflecting the chance
that $x_l < x_{(1)}$ due to the finite sample size.

\begin{figure}[h]
\plottwo{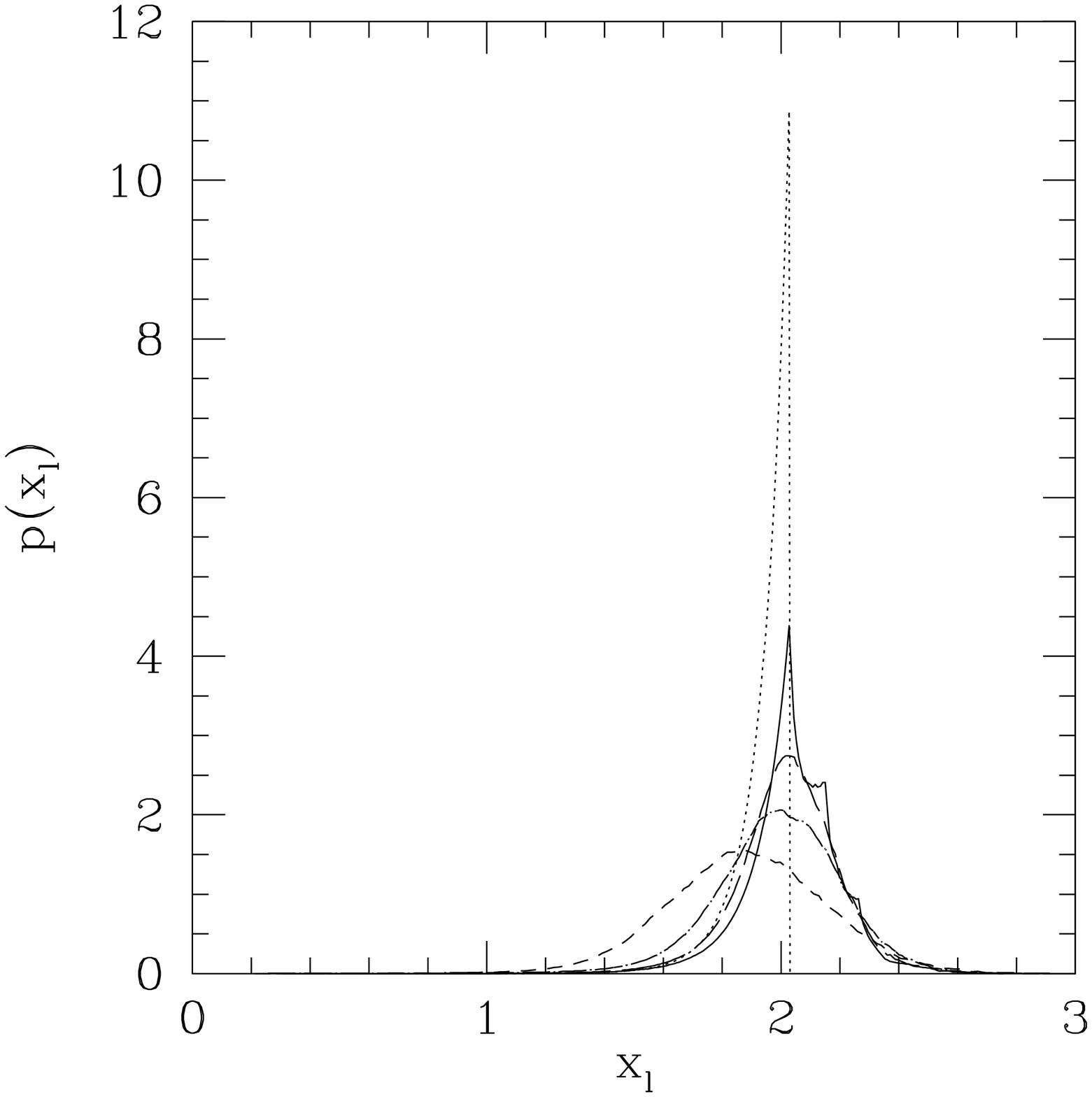}{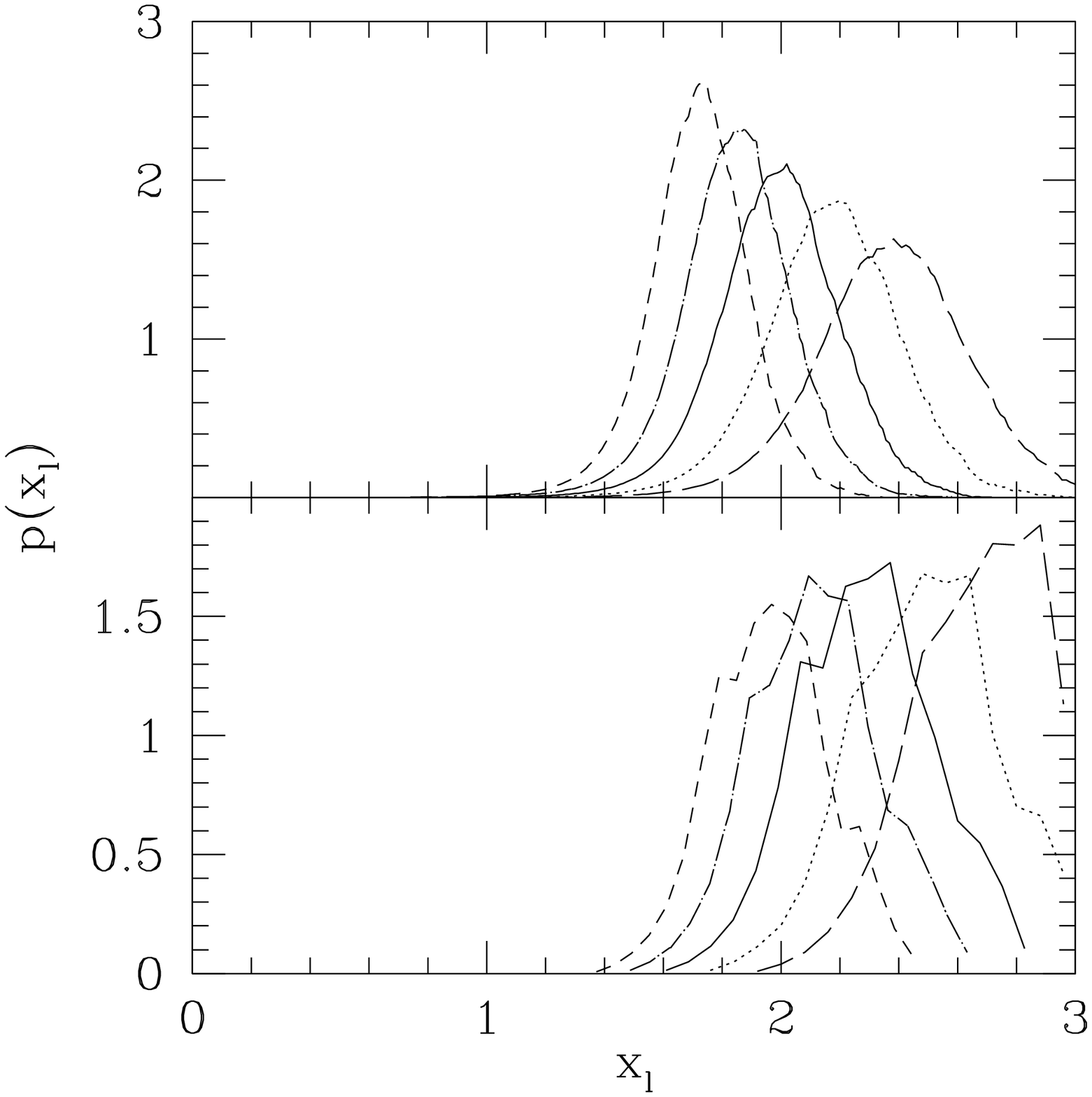}
\caption{
Left: Marginal posterior distribution for $x_l$, the location of the
inner edge for the population of hot Jupiters.  The dotted line
assumes all systems are viewed edge on ($\sin i=1$).  The solid line
assumes an isotropic distribution of orbital inclinations (except for
transiting planets).  The remaining lines replace the assumption of
$R=1.2 R_J$ for all non-transiting planets with a normal distribution
for the planet radii using a dispersion dispersion of $\sigma_{R_P} =
0.05 R_J$ (long dashes), $0.1 R_J$ (dots-dashes), or $0.2 R_J$ (short
dashes).
Upper Right: Dependence of the marginal posterior distribution for
$x_l$ on the assumed mean planet radius.  $\left<R_P\right> =
1.0 R_J$ (long dashes), $1.1 R_J$ (dotted), $1.2 R_J$ (solid), $1.3
R_J$ (dotted dashed), and $1.4 R_J$ (short dashes), all assuming
$\sigma_{R_P} = 0.1 R_J$.
Lower Right:  Same as above, but using a 2-d model (period \& mass) which accounts
for selection effects.
}
\end{figure}


Next, we assume that the actual distribution of orbital inclinations
is isotropic ($\cos i \sim U[-1,1]$).  For planets which were
discovered by radial velocities and the inclination was subsequently
determined with the detection of transits, we use the measured
inclination.  The marginal posterior distribution for $x_l$ is shown
in Fig.\ 2 (left, solid line).  The sharp cutoff at $x_{(1)}$ is replaced
with a more gradual tail, reflecting the chance that $\sin i < 1$ for
planets with the smallest values of $x$.  


Next, we consider the consequences of allowing for a distribution of
planetary radii.  For transiting planets we use a normal distribution
for the radius based on the published radius and uncertainty. For
nontransitting planets, we assume a normal distribution of planetary
radii with standard deviation, $\sigma_{R_P}$.  We show the resulting
marginalized posterior distributions in Fig.\ 2 (left).  Allowing for a
significant dispersion broadens the posterior distribution for $x_l$
and results in a slight shift to smaller values.


We have also explored the effects of varying the model parameter
$P_{\max}$, exploring values from 8d to 60d.  We find that this does not
make a discernible difference in the posterior distribution for $x_l$.


Our results are sensitive to our choice for the mean radius for the
non-transiting planets.  In Fig.\ 2 (upper right) we show the posterior
distributions for various mean radii, assuming $\sigma_{R_P}
= 0.1 R_J$.  Since few planets have a known inclination, there is
be a nearly perfect degeneracy between $R_P$ and $x_l$.  Even when
we include transiting planets, this degeneracy remains near perfect,
i.e., $p(x_l | R_p, x_1, ..., x_n ) \simeq p(x_l \cdot \frac{R'_P}{R_P} | R'_P, x_1, ..., x_n)$.
However, it can be seen that is extremely
unlikely for $x_l$ to be near unity for any reasonable planetary
radius.


We have also performed an improved analysis using a two dimensional
model which considers the joint planet mass-period distribution
function.  This allows us to account for observational selection
biases due to the minimum mass for detecting a planet depending on the
orbital period.  We assume that the distribution function is a
truncated power law in both planet-star mass ratio and period.  That
is
\be
p(P, \mu | \alpha, \beta, P_{\min} P_{\max}, \mu_{\min}, \mu_{\max}, c ) \sim c P^\alpha \mu^\beta \frac{dP}{P} \frac{d\mu}{mu},
\ee
provided $\mu_{\min} < \mu < \mu_{\max}$, $P < P_{\max}$, and $a(P,
M_*) \ge x_l \cdot a_R(R_P, \mu)$.  Here $\mu = M_P / M_*$, and
$\alpha$ and $\beta$ are the new power law indices.  We find the
marginal posterior distribution for $x_l$ is very
similar to the results of our 1-d analysis.  The most significant difference
is that the posterior distribution for $x_l$ shifts slightly to wards
{\em larger} separations (see Fig.\ 2, lower right).

\section{Discussion}

The current distribution of hot Jupiters discovered by radial velocity
searches shows a cutoff that is a function of orbital period and
planet mass.  Our Bayesian analysis solidly rejects the hypothesis
that the cutoff occurs inside or at the Roche limit, in contrast to
what would be expected if the hot Jupiters had slowly migrated inwards
on a circular orbit.  Instead, our analysis shows that this cutoff
occurs at a distance nearly twice that of the Roche limit, as expected
if the hot Jupiters were circularized from a highly eccentric orbit.
These findings suggest that hot Jupiters may have formed via
planet-planet scattering (e.g., Rasio \& Ford 1996), tidal capture of
free floating planets (Gaudi 2003), or secular perturbations from a
highly inclined binary companion (e.g., Holman, Touma \& Tremaine
1997).  If the hot Jupiters indeed were circularized from a high
eccentricity orbit, then this raises the challenge of explaining the
origin of giant planets with orbital periods $\sim10-100$ days.

An alternative explanation is that the planets migrated inwards on a
nearly circular orbit at a time when the planets were roughly twice
their current radii.  Future observations of low mass planets may make
it possible to test this alternative, assuming that the time evolution of their
contraction is significantly different than for Jupiter-mass planets.
A third alternative is that short-period giant planets are destroyed
by another process before they reach the Roche limit.  HST
observations of HD 209458 indicate absorption by matter presently
beyond the Roche lobe of the planet and have been interpreted as evidence
for a wind leaving the planet powered by stellar radiation (Vidal-Madjar et
al.\ 2003, 2004).  Further theoretical work will help determine under
what conditions these processes can cause significant mass loss.

Future planet discoveries will either tighten the
constraints on the model parameters or provide evidence for the
existence of planets definitely closer than twice the Roche limit.  In
particular, new discoveries of very low mass planets could better
constrain the shape of the inner cutoff as a function of mass.  In the
future, an improved analysis could also include such low-mass planets
where surveys are not yet complete.  For future theoretical work, we
hope to explore the possibility of orbital circularization occurring
at larger orbital radii, possibly in a protoplanetary disk or while
the star is young and still contracting.

\acknowledgments{ We thank Eugene Chiang, Tom Loredo, Norm Murray,
Ruth Murray-Clay, John Papaloizou, Frederic Pont for their comments.
This research was supported by NSF grants AST-0206182 and AST0507727
at Northwester and a Miller Research Fellowship at UC Berkeley.
}

\end{document}